# Helicity Spectra and Dissipation

O.G. Chkhetiani[1], E. Golbraikh[2]

[1]Space Research Institute, Russian Academy of Sciences, Profsoyuznaya str., 84/32 Moscow 117810, Russia

[2] Center for MHD Studies, Ben-Gurion University of the Negev, POB 653, Beer-Sheva 84105, Israel

Abstract

Both dissipation of helicity and it spectrum we are study on the basis of asymptotic model. Introduction into model dependence of angle between turbulent components vorticity and velocity on the governing parameters leads to the spectra of helicity observed experimentally and obtained by DNS model computations by various authors. At the same time, it settles the problem of helicity dissipation divergence with growing Reynolds number.

Recently, the issue of helicity dissipation in a turbulent flow in the presence of nonzero mean helicity has been broadly discussed by many authors (see [1, 3-5, 8, 10] and references therein). In [4], one of the first papers in this area, an expression for the boundary scale of helicity dissipation $l_\eta \sim \left(v^3 \varepsilon^2 / \eta^3\right)^{1/7}$ was obtained, which differs from Kolmogorov's scale $l_v \sim \left(v^3 / \varepsilon\right)^{1/4}$ defining the dissipative energy scale (here $v$ is kinematic viscosity and $\varepsilon$ and $\eta$ – averaged energy and helicity fluxes over the scales).

However, it is noteworthy that when deriving the expression for $l_\eta$, the authors [4] used an expression for the third moments of the velocity correlator, which is valid only within the inertial interval. Hence, the form of the derived characteristic scale of helicity dissipation should be different, and the inferences of the mentioned paper should be revised. In fact, it has been demonstrated in more recent papers (see, e.g., [1, 3, 5]) that within the framework of standard approaches, the scales of helicity and energy dissipation apparently coincide.

In a general case, we can write integral energy and helicity dissipations $D_E$ and $D_H$ as follows:

$$D_X = 2\nu \int_0^{l_\nu^{-1}} X(k) k^2 dk \qquad (1)$$

where $X = E$ or $H$, spectral density of energy or helicity, respectively. However, there remains a problem of integral helicity dissipation, since the latter increases with growing Reynolds number (at $\nu \to 0$). Really, when expanding the turbulent velocity component in helical waves [3, 9, 12], the values of spectral helicity and energy densities $H(k)$ and $E(k)$ can be presented as

$$E(k) = E^+(k) + E^-(k)$$

$$H(k) = 2k(E^+(k) - E^-(k)),$$

where $E^\pm(k)$ are spectral positively determined energy densities of helical waves. If $E(k) \sim k^{-5/3}$, and if we assume, as in [4], that $E^\pm(k)$ have a similar spectrum, the integral helicity dissipation

$$D_H = 2\nu \int_0^{l_\nu^{-1}} (E^+(k) - E^-(k)) k^3 dk \sim \nu \varepsilon^{2/3} l_\nu^{-7/3} \propto \nu^{-3/4} \qquad (2)$$

diverges at $\nu \to 0$.

If, however, we assume that the spectral helicity density obeys the same dependence on the wave vector as spectral energy density [12], i.e. $H(k) \sim \eta \varepsilon^{-1/3} k^{-5/3}$, then the integral (1) does not diverge. Hence the problem of divergence is due to the fact that we are unaware of the form of function $H(k)$ in the inertial interval.

We examine this issue from a somewhat different standpoint. It follows from quasi-Kolmogorov's approximation that in the stationary case helicity production equals its dissipation $\eta$. Then a question arises – which form of spectral density $H(k)$ this condition leads to and whether it agrees with experimental data and numerical modeling data.

In all the above-mentioned papers, helicity and energy fluxes over the spectrum were assumed, either explicitly or implicitly, to be constant not only in the inertial, but also in the dissipative range. In the asymptotic approximation [6], the spectral index $\delta_E$ appearing in the expression of the energy $E \sim r^{\delta_E}$ is connected with the parameter $\alpha_1$ (which determines the dependence of various turbulent characteristics of the flow in the dissipative range on $\nu$, $\varepsilon$ and $\eta$) as follows:

$$\delta_E = \frac{6\alpha_1 + 4}{3(1-\alpha_1)} \qquad (3)$$

An attempt to study the asymptotic behavior of the helicity spectrum using an approach similar to that suggested in [6] was made in [7]. However, when studying the behavior of the structural function of helicity, it was not taken into account that the angle $\varphi$ between the velocity and vorticity is also a random value depending on the same governing parameters ($\nu$, $\varepsilon$ and $\eta$). Taking this into account along with the fact that the characteristic value $\cos\varphi$ is dimensionless, we can write within the frames of asymptotic approach:

$$\cos\varphi \sim \varepsilon^{\alpha_2} \eta^{-\frac{4}{5}\alpha_2} \nu^{-\frac{3}{5}\alpha_2} \qquad (4)$$

where $\alpha_2$ is a certain parameter. Using [7], we obtain the following expression for the spectral index $\delta_H$ of helicity ($H(r) \sim r^{\delta_H}$):

$$\delta_H = \frac{9\alpha_1 + 1 - 3\alpha_2}{3(1-\alpha_1)} \qquad (5)$$

in contrast to $\delta_H = \frac{9\alpha_1 + 1}{3(1-\alpha_1)}$ derived in [7]. According to Eq. (4), the limiting case examined in [7] at $\cos\varphi = 1$ connects independent governing parameters $\nu$, $\varepsilon$ and $\eta$ and contradicts the assumption of their independence in the general case accepted in this paper.

As follows from the definition of the value $\eta$ in Navier-Stokes equation,

$$\eta = \nu \langle u' \cdot \Delta \mathrm{curl} u' \rangle \qquad (6)$$

where $u'$ is a turbulent component of the velocity field and $\langle ... \rangle$ denotes averaging over the ensemble.

Thus, to the first approximation, we can write

$$\eta \sim \nu \frac{U_0 W_0 \cos\varphi}{l_0^2} \qquad (7)$$

where the characteristic values of the velocity $U_0 \sim \varepsilon^{-\alpha_1}\eta^{1/5(1+4\alpha_1)}v^{1/5(2+3\alpha_1)}$, vorticity $W_0 \sim \varepsilon^{-2\alpha_1}\eta^{2/5(1+4\alpha_1)}v^{1/5(6\alpha_1-1)}$ and scale $l_0 \sim \varepsilon^{\alpha_1}\eta^{-1/5(1+4\alpha_1)}v^{3/5(1-\alpha_1)}$ are determined in compliance with [6, 7]. Taking (4) into account, the right-hand side of (7) acquires the form:

$$\varepsilon^{-5\alpha_1+\alpha_2}\eta^{(1+4\alpha_1)-4/5\alpha_2}v^{3\alpha_1-3/5\alpha_2} \tag{8}$$

Since (7) should hold at any $\alpha_1$ value, following (8) we obtain that $\alpha_2 = 5\alpha_1$, and the spectral index of helicity according to (5) is:

$$\delta_H = \frac{1-6\alpha_1}{3(1-\alpha_1)} \tag{9}$$

It follows from expressions (3) and (9) that spectral indices of energy and helicity coincide at $\alpha_1 = -\frac{1}{4}$ ($\delta_E = \delta_H = \frac{2}{3}$), i.e. in the Kolmogorov's case [6]. However, in the helical case at $\alpha_1 = 0$, $\delta_E = \frac{4}{3}$ and $\delta_H = \frac{1}{3}$.

As follows from (4), $\cos\varphi \sim \varepsilon^{5\alpha_1}\eta^{-4\alpha_1}v^{-3\alpha_1}$. In Kolmogorov's case, $\cos\varphi \sim \varepsilon^{-5/4}\eta\, v^{3/4}$ and the effective angle between the velocity and vorticity increase (helicity decreases) with growing Reynolds parameter ($v \to 0$), and turbulence has only dissipative properties (Lamb's vector grows). On the other hand, helicity is maximal in the helical case ($\cos\varphi \approx const$), and its dependence on the governing parameters is minimal. Generation properties of this kind of turbulence are maximal (see [2] and references therein).

Thus, the account for the dependence of fluctuations of the angle between the velocity and vorticity on governing parameters of the flow has enabled us to obtain the spectral behavior of helicity observed experimentally [11] and obtained by model computations by various authors [1, 3, 9, 10] within the frames of an asymptotic model. It is noteworthy that the spectral index of helicity is connected with the spectral index of energy according to (3) and (9), which allows us to determine one of them through another. Here the Kolmogorov case is special, since in this case the indices are equal. At the same time, it settles the problem of helicity dissipation divergence with growing Reynolds number, since it is assumed to be initially constant ($\sim \eta$), and the arising helicity spectra are rigidly connected with its constancy.

E.G. pleasure to thank Prof. A. Tsinober for useful discussions. Work was partially supported of O.C. by RFBR grant 05-05-64735